# NANOSTRUCTURED IMMUNOSENSORS. APPLICATION TO THE DETECTION OF PROGESTERONE

*J.M. Pingarrón, P. Yáñez-Sedeño, A. González-Cortés, V. Carralero*

Department of Analytical Chemistry. Faculty of Chemistry. University Complutense of Madrid.
28040- Madrid. Spain

## ABSTRACT

A novel nanostructured electrochemical immunsensor for the determination of progesterone is reported. The approach combines the properties of gold nanoparticles with the use of a graphite-Teflon composite electrode matrix, into which gold nanoparticles are incorporated by simple physical inclusion. The antibody anti-progesterone was directly attached to the electrode surface. The immunosensor functioning is based on competitive assay between progesterone and alkaline phosphatase-labelled progesterone. Monitoring of the affinity reaction was accomplished by the electrochemical oxidation of 1-naphtol. Modification of the graphite -Teflon electrode matrix with gold nanoparticles improves substantially the electrooxidation response of 1-naphtol. Using a detection potential of +0.3V, a detection limit for progesterone of 0.84 ng ml$^{-1}$ was obtained. Analysis of seven milk samples spiked at a 3.5 ng ml$^{-1}$ progesterone concentration level yielded a mean recovery of 101$\pm$6%. Detection of the antigen-antibody reaction with a graphite - Teflon - colloidal - gold - Tyrosinase electrode, using phenylphosphate as alkaline phosphatase substrate to generate phenol, which is subsequently reduced at -0.1 V at the composite electrode, produced a high improvement in the sensitivity for progesterone detection

## 1. INTRODUCTION

The use of nanoparticles for the construction of biosensing devices is nowadays one of the most exciting trends in modern bioanalytical chemistry. Of course, one of the most important goals in the preparation of reliable biosensors is to achieve immobilization of biological recognition elements without significant losses of their activity. In particular, the use of gold nanoparticles allows the adsorption of biological molecules retaining their biological activity [1]. Considering specifically electrochemical biosensors, modification of electrodes with gold nanoparticles provides a microenvironment similar to that of proteins in native systems, with gold nanocrystals acting as conducting tunnels and facilitating electron transfer [2].

In this communication, a novel electrochemical immunosensor for the determination of progesterone is proposed. The approach combines the properties of gold nanoparticles with the use of a graphite - Teflon composite electrode matrix, into which gold nanoparticles are incorporated by simple physical inclusion. Moreover, the experimental approach implied the direct attachment of the antibody anti-progesterone to the electrode surface.

## 2. EXPERIMENTAL

1. *Fabrication of the gold nanoparticle-graphite-Teflon composite electrode*

A colloidal gold suspension, with gold nanoparticles diameter of 16$\pm$2 nm, was prepared as described previously [3]. Then, 150 mg of graphite powder (Ultra Carbon, Bay City, MI, USA), and 900 µl of the colloidal gold suspension were thoroughly mixed for 2h. Water was evaporated under air current, and then 450 mg of Teflon powder (Aldrich) were added and thoroughly mixed by hand. The mixture was pressed into pellets at 10,000 Kg cm$^{-2}$ for 10 min. The resulting pellets were 1.3 cm diameter and around 0.4 cm thick. From each of these pellets, five or six 3.0 mm diameter cylindrical portions were bored, each portion constituting a different composite electrode. Each electrode was press-fitted into a Teflon holder, and the electrical contact was made through a stainless steel screw.

2. *Fabrication of the immunosensor*

5 µl of a 0.5 µg ml$^{-1}$ monoclonal antiprogesterone (anti-Prog) (Sigma) solution was deposited on the gold nanoparticles-modified graphite-Teflon electrode surface and allowed to dry at air under room temperature. Then, the bioelectrode was immersed for 5 min into a 2 % (w/w) BSA solution prepared in 0.1 M diethanolamine-HCl buffer of pH 7.2, and washed carefully with a DEA buffer solution of pH 10.0.

3. *Electrochemical detection of progesterone*

The progesterone detection was accomplished by depositing 8 µl of the appropriate 5:3 (v/v) alkaline phosphatase-labelled progesterone (Prog-AP, Ridgeway Science Ltd, Alvington, Gloucestershire, UK): progesterone (Prog, Aldrich) solution, on the immunosensor surface. After drying at room temperature, the immunosensor with the immuno-conjugate layer was placed into the electrochemical cell containing 1 mM $MgCl_2$ and 0.1 M diethanolamine-HCl buffer solution of pH 10.0. Next, 500 µl of 0.1 M 1-naphtyl phosphate (Fluka) were added, and the corresponding oxidation current from 1-naphtol was amperometrically measured at +0.3 V vs. Ag/AgCl.

## 3. RESULTS AND DISCUSSION

The immunosensor functioning is depicted in Scheme 1, where it can be observed as it is based on competitive assay between Prog and Prog-AP. Monitoring of the affinity reaction was carried out by the electrochemical oxidation of 1-naphtol as the product of the alkaline phosphatase enzyme reaction with 1-naphtylphosphate used as the enzyme substrate.

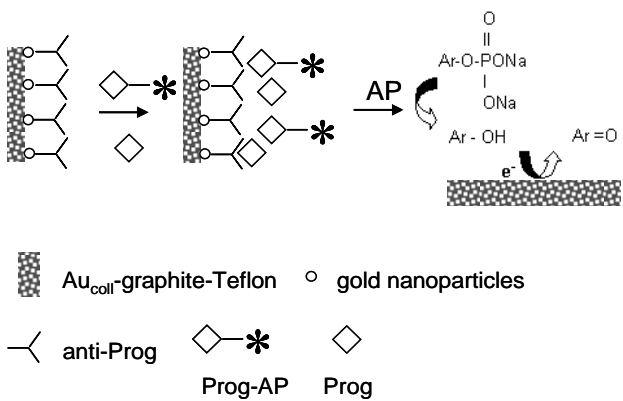

Scheme 1

Modification of the graphite-Teflon electrode matrix with gold nanoparticles improves substantially the electro-oxidation response of 1-naphtol. So, Figure 1 shows cyclic voltammograms recorded at both the modified and the unmodified electrode. As can be observed, the peak current increased remarkably at the gold nanoparticles - modified electrode as a consequence of the enhanced electron transfer kinetics in the presence of gold nanoparticles.

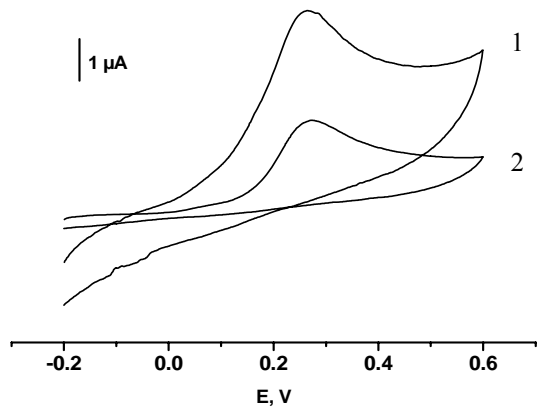

Figure 1. Cyclic voltammograms of $5.0 \times 10^{-4}$ M 1-naphtol at (1) $Au_{coll}$-graphite-Teflon and (2) graphite-Teflon electrodes. Supporting electrolyte, 0.1 mol $l^{-1}$ diethanolamine-HCl buffer solution of pH 10.0.

Using a detection potential of +0.3 V vs. Ag/AgCl, the steady-state current measured for $10^{-4}$ M 1-naphtol was 43% higher at the gold nanoparticles-modified electrode that at the unmodified graphite-Teflon electrode.

Different experimental variables involved both on the immunosensor preparation and the immunosensing conditions were optimized. The selected parameters are summarized in Table 1.

Table 1. Experimental variables selected for the construction of a progesterone biosensor based on gold nanoparticles -graphite-Teflon electrodes

| | |
|---|---|
| Teflon percentage | 70 % |
| Colloidal gold | 900 µl |
| Amount of antibody | 0.5 µg $ml^{-1}$ |
| 1-naphtylphosphate concentration | 5 mM |
| pH | 10.0 |
| Prog-AP/Prog ratio | 5:3 |
| Prog-AP/Prog volume | 8 µl |

Under these experimental conditions, as expected, the steady-state current decreased as the concentration of progesterone increased, and this decrease is related to the antigen concentration. Furthermore, the analyte responses were fast, the steady-state current being reached in 30 - 40s from the addition of the enzyme substrate. A linear range (r = 0.991) for progesterone was found over the 0 - 30 ng ml$^{-1}$ concentration range, which is useful to analyze progesterone in milk, with a slope value of -1.2 nA ng$^{-1}$ ml. It should be pointed out that this slope value is approximately 8-fold higher than that reported for screen-printed carbon electrode progesterone biosensors coated with IgG [4]. The limit of detection according to the ΔS/N = 3 criterion was 0.84 ng ml$^{-1}$.

A RSD value of 3.5 % was obtained for 10 successive amperometric responses from a 20 ng ml$^{-1}$ progesterone solution, indicating a good repeatability of the electroanalytical measurements with the same immunosensor. Furthermore, responses obtained with three different immunosensors yielded a RSD value of 8.2 %, showing an acceptable reproducibility in the biosensor construction.

The developed immunosensor was applied to the determination of progesterone in spiked milk samples at a 3.5 ng ml$^{-1}$ progesterone concentration level. The methodology employed was extremely simple since it involved only dilution of spiked milk with the diethanolamine - HCl buffer solution of pH 10. A 3-μl aliquot of this solution were mixed with 5 μl of Prog-AP solution and then deposited on the electrode surface.

A calibration curve for progesterone following this procedure was needed to be constructed as a consequence of the remarkable matrix effect observed. Using this calibration curve, the analysis of seven milk samples yielded a mean progesterone concentration value of 3.5 $\pm$ 0.2 ng ml$^{-1}$ (mean recovery of 101 $\pm$ 6%), which demonstrated fairly well the usefulness of the immunosensor for this type of analysis.

*Detection at a graphite - Teflon - gold nanoparticles-Tyrosinase electrode*

An alternative detection method of the antigen-antibody reaction based on the use of a composite graphite-Teflon-colloidal gold - Tyrosinase bioelectrode has been also evaluated. This biosensor exhibits suitable amperometric responses at -0.10 V for different phenolic compounds [5]. The presence of colloidal gold into the composite matrix was demonstrated to enhance kinetics of both the enzyme reaction involving Tyrosinase, and the electrochemical reduction of the corresponding o-quinones at the electrode surface. Therefore, these advantages were profited to monitor the anti-Prog-Prog affinity reaction.

The antibody was also directly attached onto the electrode surface and the immunosensor is also based on competitive assay between Prog and Prog-AP, but, in this case the AP- substrate was phenylphosphate in order to generate phenol, which is subsequently reduced at -0.1 V at the Tyrosinase-colloidal gold composite electrode. A scheme of the biosensor functioning is depicted in Scheme 2.

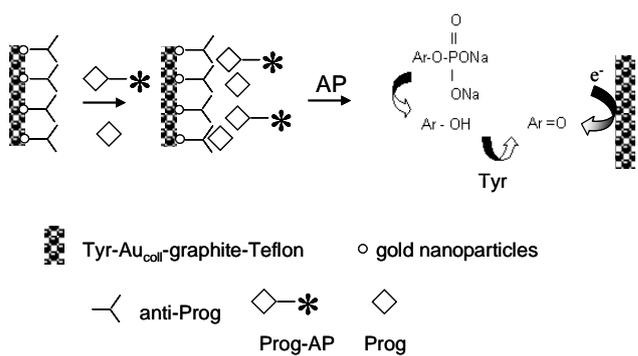

Scheme 2

Apart from the fact of measuring at a potential such as -0.1 V, the enhancement of the reaction kinetics due to the presence of colloidal gold allows the achievement of a high sensitivity for the detection of phenol. This produces a further improvement in the sensitivity for the progesterone detection. So, a linear calibration curve for progesterone (r=0.995) was constructed in the 0 - 50 ng ml$^{-1}$ concentration range, with a slope value of -82.3 nA ng$^{-1}$ ml. As can be observed, the sensitivity achieved now is almost one hundred-fold higher than that provided with the former detection approach, and consequently, almost one thousand higher than that reported with screen-printed progesterone biosensors coated with IgG. The limit of detection, calculated for a signal-to-noise ratio equal to 3, was 0.34 ng ml$^{-1}$. Moreover, 10 successive measurements for a 10 ng ml$^{-1}$ progesterone concentration yielded a RSD value of 4 %. Amperometric responses obtained with six different immunosensors at the same progesterone concentration level gave a RSD value of 2.3 %, which demonstrated the excellent reproducibility achieved in the construction of the nanostructured biosensors.

## 3. CONCLUSIONS

The sinergic combination of the advantages provided by nanostructured biosensors with gold nanoparticles and the use of a composite bioelectrode matrix, into which the nanoparticles are incorporated by physical inclusion, permits the development of robust immunosensors in which the antibody is directly attached to the electrode surface. The detection approaches allowed by this immunosensor design, imply a significant improvement of the analytical characteristics for the determination of progesterone with respect to previous immunosensor designs, and permits their application for the analysis of real samples.